\documentclass[aps,manuscript,pop,showpacs,preprint,superscriptaddress]{revtex4-1}
\usepackage{lmodern}

\usepackage{lmodern}
\usepackage[T1]{fontenc}
\usepackage[latin9]{inputenc}
\usepackage{amsthm}
\usepackage{amsmath}
\usepackage{graphicx}

\makeatletter

\@ifundefined{textcolor}{}
{%
 \definecolor{BLACK}{gray}{0}
 \definecolor{WHITE}{gray}{1}
 \definecolor{RED}{rgb}{1,0,0}
 \definecolor{GREEN}{rgb}{0,1,0}
 \definecolor{BLUE}{rgb}{0,0,1}
 \definecolor{CYAN}{cmyk}{1,0,0,0}
 \definecolor{MAGENTA}{cmyk}{0,1,0,0}
 \definecolor{YELLOW}{cmyk}{0,0,1,0}
}
\theoremstyle{plain}
\newtheorem{thm}{\protect\theoremname}

\usepackage{amsthm}\usepackage{latexsym}\usepackage{bm}\usepackage{amsfonts}

\makeatother

\providecommand{\theoremname}{Theorem}

\begin{document}

\title{On the structure of the two-stream instability -- complex G-Hamiltonian
structure and Krein collisions between positive- and negative-action
modes}

\author{Ruili Zhang}

\affiliation{Department of Modern Physics and School of Nuclear Science and Technology,
University of Science and Technology of China, Hefei, Anhui 230026,
China}

\affiliation{Key Laboratory of Geospace Environment, CAS, Hefei, Anhui 230026,
China}

\author{Hong Qin }

\thanks{Corresponding author, hongqin@princeton.edu}

\affiliation{Plasma Physics Laboratory, Princeton University, Princeton, NJ 08543,
USA}

\affiliation{Department of Modern Physics and School of Nuclear Science and Technology,
University of Science and Technology of China, Hefei, Anhui 230026,
China}

\author{Ronald C. Davidson}

\affiliation{Plasma Physics Laboratory, Princeton University, Princeton, NJ 08543,
USA}

\author{Jian Liu}

\affiliation{Department of Modern Physics and School of Nuclear Science and Technology,
University of Science and Technology of China, Hefei, Anhui 230026,
China}

\affiliation{Key Laboratory of Geospace Environment, CAS, Hefei, Anhui 230026,
China}

\author{Jianyuan Xiao}

\affiliation{Department of Modern Physics and School of Nuclear Science and Technology,
University of Science and Technology of China, Hefei, Anhui 230026,
China}

\affiliation{Key Laboratory of Geospace Environment, CAS, Hefei, Anhui 230026,
China}
\begin{abstract}
The two-stream instability is probably the most important elementary
example of collective instabilities in plasma physics and beam-plasma
systems. For a warm plasma with two charged particle species based
on a 1D warm-fluid model, the instability diagram of the two-stream
instability exhibits an interesting band structure that has not been
explained. We show that the band structure for this instability is
the consequence of the Hamiltonian nature of the warm two-fluid system.
Interestingly, the Hamiltonian nature manifests as a complex G-Hamiltonian
structure in wave-number space, which directly determines the instability
diagram. Specifically, it is shown that the boundaries between the
stable and unstable regions are locations for Krein collisions between
eigenmodes with different Krein signatures. In terms of physics, this
rigorously implies that the system is destabilized when a positive-action
mode resonates with a negative-action mode, and that this is the only
mechanism by which the system can be destabilized. It is anticipated
that this physical mechanism of destabilization is valid for other
collective instabilities in conservative systems in plasma physics,
accelerator physics, and fluid dynamics systems, which admit infinite-dimensional
Hamiltonian structures.
\end{abstract}
\maketitle

\section{Introduction \label{sec:Introduction}}

The two-stream instability is a fundamental collective instability
in plasma physics. In addition to its practical importance in many
application areas \cite{Davidson99,1davidson2001physics,qinhong2000,zimmermann2004,qin2014two},
it has been studied as one of the most elementary example of plasma
instabilities \cite{Kueny95,Morrison14,Lashmore07}. For a homogeneous
cold plasma with two charged particle species drifting in the $z$-direction
with different macroscopic velocities, the 1D linear electrostatic
perturbations of the form $\sim\exp(ikz-i\omega t)$ are unstable
(exhibit exponential temporal growth) when the following well-known
condition is satisfied \cite{1davidson2001physics},
\begin{equation}
0<k^{2}V^{2}<(\omega_{p1}^{2/3}+\omega_{p2}^{2/3})^{3}\,,
\end{equation}
where $\omega_{pj}\thinspace(j=1,2)$ is the plasma frequency, $v_{j}^{0}\thinspace(j=1,2)$
is the drift velocity of the $j$-th species, and $V=v_{2}^{0}-v_{1}^{0}$
is the relative drift velocity. Note that $(\omega_{p1}^{2/3}+\omega_{p2}^{2/3})^{3/2}/k$
is the threshold value for the relative velocity, above which the
mode is stabilized.

\begin{figure}
\includegraphics[scale=0.6]{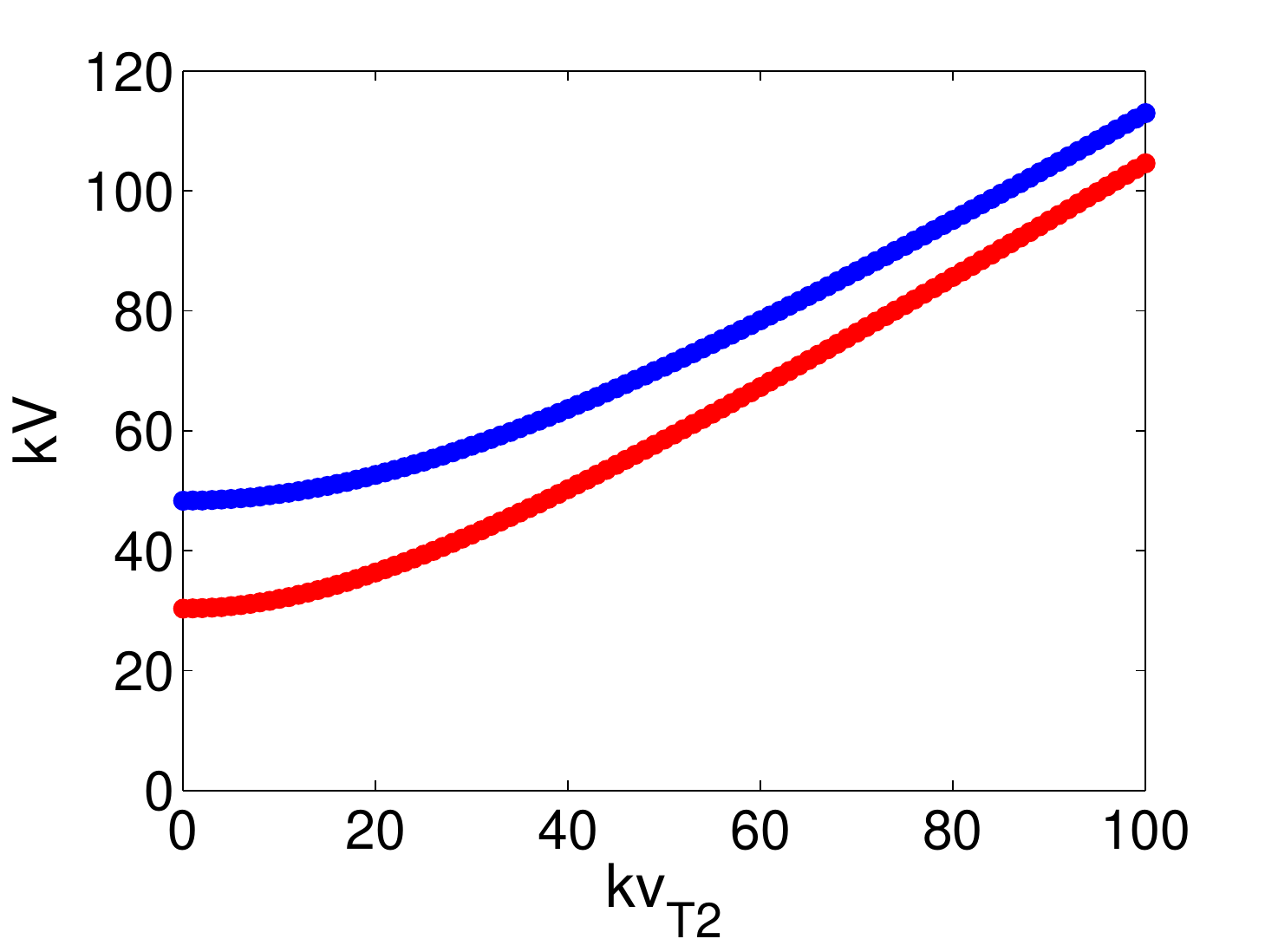}\protect\caption{Instability diagram for the warm two-stream instability in the parameter
plane of $kV$ and $kv_{T2}$. Other system parameters are chosen
to be $\omega_{p1}^{2}=1,\thinspace\omega_{p2}^{2}=1836,\thinspace kv_{1}^{0}=0,$
and $kv_{T1}=1.$ The connected region between the upper curve and
the lower curve is the unstable band, and the other two disconnected
regions are the stable regions. \label{f1}}
\end{figure}

In this paper, we study the stability properties of the two-stream
instability when the thermal velocities $v_{Tj}\thinspace(j=1,2)$
of the two species are non-vanishing. The dispersion relation for
two-stream excitations is given by \cite{qin2014two}
\begin{equation}
\dfrac{\omega_{p_{1}}^{2}}{(\omega-kv_{1}^{0})^{2}-k^{2}v_{T1}^{2}}+\dfrac{\omega_{p_{2}}^{2}}{(\omega-kv_{2}^{0})^{2}-k^{2}v_{T2}^{2}}=1\thinspace,\label{eq:DR}
\end{equation}
which is derived from the warm two-fluid model in Sec. \ref{sec:model}.
In the present study, we normalize $(\omega,\thinspace\omega_{pj},\thinspace kv_{j}^{0},\thinspace kV,\thinspace kv_{Tj})$
by $\omega_{p1}$. In principle, the dispersion relation in Eq.\,\eqref{eq:DR}
can be solved analytically. But it is straightforward to solve it
numerically. One example is given in Fig.\,\ref{f1}, where the instability
diagram is presented in the parameter plane corresponding to $kV$
and $kv_{T2}$, and other parameters are chosen to be $\omega_{p1}^{2}=1,\thinspace\omega_{p2}^{2}=1836,\thinspace kv_{1}^{0}=0,$
and $kv_{T1}=1.$ In Fig.\,\ref{f1}, the connected region between
the upper curve and the lower curve is the unstable band, and the
other two disconnected regions are the stable regions.

Comparing with the cold two-stream instability, the instability diagram
for the warm two-stream instability is much more interesting. The
unstable region is a connected band, the lower threshold for the relative
velocity $V$ is larger than zero, and there is no upper threshold
for the relative velocity $V$. This is drastically different from
the cold two-stream instability, which has an upper threshold $(\omega_{p1}^{2/3}+\omega_{p2}^{2/3})^{3/2}/k$
for the relative velocity $V.$ For the warm two-stream instability,
for any value of $V$, about the lower threshold there is always a
thermal velocity $v_{T2}$ that destabilizes the mode. For a fixed
relative velocity $V$, as $v_{T2}$ increases from $0$ to large
values, the mode is at first stable, and then become unstable at a
lower critical value, after which the mode is unstable until $v_{T2}$
reaches an upper critical value. The modes becomes stable again after
$v_{T2}$ increases beyond the upper threshold. We observe a similar
behavior when $V$ varies for a fixed value of $v_{T2}.$

The purpose of the present study is to reveal the physical and mathematical
structure of the band structure of the warm two-stream instability.
It turns out, in term of the key physics, that the band structure
is largely due to the conservative Hamiltonian nature of the two-stream
interaction. And the mathematical description of the Hamiltonian nature
is expressed as a complex G-Hamiltonian structure, which is a complex,
non-canonical generalization of the familiar real canonical Hamiltonian
structure. The mathematical theory of the linear G-Hamiltonian system
has been systematically developed by Krein, Gel'fand and Lidskii \cite{Krein1950,Gel1955,KGML1958}.
As the system parameters vary, a necessary and sufficient condition
for the onset of instability is that two eigenmodes with opposite
Krein signatures collide (see Sec.\,\ref{sec:PB}). This is the so-called
Krein collision. We will show that the physical meaning of the Krein
signature is the sign of the action for the eigenmode. For a stable
mode, the action is defined to be the ratio between the energy and
the frequency of the eigenmode. We will show that the instability
boundaries in Fig.\,\ref{f1} are the locations for the Krein collisions.
In terms of essential physics, the instability boundaries are located
where a positive-action mode resonates with a negative-action mode.

To the best of our knowledge, the study present in this paper is the
first of its kind to systematically apply the G-Hamiltonian theoretical
framework to plasma physics applications. However, we would like to
mention a few similar concepts and methods that have been discussed
by the plasma physics community. The first concept is the Hamiltonian-Hopf
bifurcation. To those who are familiar with Hamiltonian dynamics,
the boundary between a stable and unstable equilibrium is known as
the Hamiltonian-Hopf bifurcation. One may wonder whether the instability
boundaries in Fig.\,\ref{f1} are Hamiltonian-Hopf bifurcations.
The answer is no. This is because the Hamiltonian-Hopf bifurcation
is specifically for a real Hamiltonian system, and our dynamical system
in time is a complex G-Hamiltonian system for one Fourier component
in space. To a certain extent, the Krein collsion at the instability
boundaries in Fig.\,\ref{f1} can be viewed as a complex generalization
of the Hamiltonian-Hopf bifurcation. Since all dynamical systems in
plasmas are complex after Fourier decomposition in space, we expect
that this feature applies to all plasma instabilities in infinite
dimensional Hamiltonian systems, i.e., an unstable mode is created
and can only be created by the resonance between two stable modes
with opposite-sign actions.

Another related concept is that of the negative- and positive-energy
modes \cite{Sturrock58,Kueny95,Lashmore07,Morrison14}. Such concepts
have appeared from time to time in the plasma physics literature,
mostly in connection with the destabilization of certain modes. It
has been suggested that instability occurs when a negative-energy
mode resonates with a positive-energy mode. However, this process
has only been discussed at an intuitive level without a rigorous mathematical
description. What we show in the present study is that the relevant
concept is not that of negative or positive energy. Instead, what
matters is whether the action is negative or positive. We will provide
a rigorous definition of the action of an eigenmode, and use the example
of the warm two-stream instability to show mathematically that system
becomes unstable when and only when a positive-action mode resonates
with a negative-action mode.

The Hamiltonian structure of the linear two-stream equation system
for a single $k$ was first studied by Kueny and Morrison \cite{Kueny95,Morrison14}.
When the system is stable, a set of coordinate transformations were
given to transform the complex system into an 8-D real canonical normal
form, which applies to cases with two distinct real eigenfrequencies.
In the general case, the dispersion relation \eqref{eq:DR} for the
warm two-stream modes admits four distinct complex eigenfrequencies.
Particularly noteworthy, both the real and imaginary parts of an eigenfrequency
can be non-vanishing (see Figs. \ref{f2} and \ref{f3}). The complex
G-Hamiltonian structure for the warm two-stream system discovered
in the present study applies to the most general warm two-stream modes.
This should not be surprising. As a matter of fact, the only symmetry
constraint that the dispersion relation \eqref{eq:DR} imposes is
that the eigenfrequencies are symmetric with respect to the real axis,
and this is also the only symmetry constraint that a G-Hamiltonian
matrix admits (see Theorem \ref{thm:es}). This is why a $4\times4$
complex G-Hamiltonian structure is able to faithfully represent the
general spectrum of the warm two-stream system. On the other hand,
we can attempt to use a real symplectic system to represent the spectrum
of the warm two-stream system. However, we note that the eigenfrequencies
of a $sp(2n)$ matrix are symmetric with respect to both the real
and imaginary axes, and the constraint of being symmetric with respect
to the imaginary axis is not that of the warm two-fluid system. Therefore,
one has to use a real symplectic system whose dimension is at least
16. Here, $sp(2n)$ is the Lie algebra of the Lie group of symplectic
matrices $Sp(2n).$

In plasma physics and accelerator physics, there are many other applications,
e.g., charged particle dynamics in a periodic focusing lattice \cite{Qin09PRSTAB,PhysRevLett.111.104801,Moses13},
where the underlying systems are finite-dimensional real Hamiltonian
systems. In these cases, the Krein collision theory is also valid
for these real symplectic systems. Stability analyses using the Krein
collision theory have been successfully applied to these systems to
generate results of practical importance \cite{Qin14PRSTAB-044001,Qin15-056702,Groening14}
.

This paper is organized as follows. In Sec.\,\ref{sec:model}, the
dispersion relation for the warm two-stream instability is derived
from a set of linearized warm two-fluid equations. In Sec.\,\ref{sec:PB}
the theory of G-Hamiltonian systems and Krein collisions are introduced.
We then show that the warm two-stream system in wave-number sapce
is a G-Hamiltonian system in Sec.\,\ref{sec:TS}, and analyze the
structure of the instability diagram using the physical mechanism
of resonance between positive- and negative-action modes.

\section{Theoretical model for warm two-stream instability \label{sec:model}}

We consider the warm two-fluid system describing a non-relativistic
plasma with 1D spatial variations and electrostatic fields in the
z-direction,
\begin{align}
\dfrac{\partial n_{j}}{\partial t}+\dfrac{\partial}{\partial z}(n_{j}v_{j})=0,\\
\dfrac{\partial v_{j}}{\partial t}+v_{j}\dfrac{\partial v_{j}}{\partial z}+\dfrac{1}{n_{j}m_{j}}\dfrac{\partial p_{j}}{\partial z}=\dfrac{e_{j}}{m_{j}}E,\\
\dfrac{\partial E}{\partial z}=\sum_{j}4\pi e_{j}n_{j},\\
p_{j}=\dfrac{p_{j0}}{\hat{n_{j}}^{\gamma_{j}}}n_{j}^{\gamma_{j}},
\end{align}
where $j=1,2$ is the index labeling plasma species, $e_{j}$ and
$m_{j}$ are the charge and mass of a particle of species $j$, $\gamma_{j}$
is the polytropic index, and $p_{j0}$ and $\hat{n}_{j}$ are constants.
Considering small perturbations $\widetilde{\rho}_{j}=e_{j}\widetilde{n}_{j}$
and $\widetilde{v}_{j}$ about the constant values $e_{j}\hat{n}_{j}$
and $v_{j}^{0}$, we obtain the linearized fluid equations in terms
of $\widetilde{\rho}_{j}$,
\begin{align}
\left(\dfrac{\partial}{\partial t}+v_{1}^{0}\right)\left(\dfrac{\partial}{\partial t}+v_{1}^{0}\right)\widetilde{\rho}_{1}-v_{T1}^{2}\widetilde{\rho}_{1}=-\omega_{p1}^{2}(\widetilde{\rho}_{1}+\widetilde{\rho}_{2}),\label{eq:7}\\
\left(\dfrac{\partial}{\partial t}+v_{2}^{0}\right)\left(\dfrac{\partial}{\partial t}+v_{2}^{0}\right)\widetilde{\rho}_{2}-v_{T2}^{2}\widetilde{\rho}_{1}=-\omega_{p2}^{2}(\widetilde{\rho}_{1}+\widetilde{\rho}_{2}),\label{eq:8}
\end{align}
 where $\omega_{pj}^{2}=4\pi\hat{n}_{j}e_{j}^{2}/m_{j}$ and $v_{Tj}^{2}=\gamma_{j}T_{j}/m_{j}$.
For a single Fourier mode in space $\widetilde{\rho}_{j}\sim\exp(ikz)$,
where $k$ is the wavenumber of the perturbation, Eqs.\,\eqref{eq:7}
and \eqref{eq:8} reduce to two coupled ordinary differential equations
\begin{align}
\big(\frac{d}{dt}+ikv_{1}^{0}\big)^{2}\widetilde{\rho}_{1}+k^{2}v_{T1}^{2}\widetilde{\rho}_{1} & =-\omega_{p1}^{2}(\widetilde{\rho}_{1}+\widetilde{\rho}_{2}),\\
\big(\frac{d}{dt}+ikv_{2}^{0}\big)^{2}\widetilde{\rho}_{2}+k^{2}v_{T2}^{2}\widetilde{\rho}_{2} & =-\omega_{p2}^{2}(\widetilde{\rho}_{1}+\widetilde{\rho}_{2}),
\end{align}
 which can also be expressed in the compact form of a 4-dimensional
linear complex dynamical system,
\begin{equation}
\dot{\mathbf{x}}=A\mathbf{x},\label{eq:AA}
\end{equation}
where
\begin{align}
A & =\left(\begin{array}{cccc}
-2ikv_{1}^{0} & 0 & (kv_{1}^{0})^{2}-\omega_{p1}^{2}-k^{2}v_{T1}^{2} & -\omega_{p1}^{2}\\
0 & -2ikv_{2}^{0} & -\omega_{p2}^{2} & (kv_{2}^{0})^{2}-\omega_{p2}^{2}-k^{2}v_{T2}^{2}\\
1 & 0 & 0 & 0\\
0 & 1 & 0 & 0
\end{array}\right),\\
\mathbf{x} & =\left(\begin{array}{c}
d\widetilde{\rho}_{1}/dt\\
d\widetilde{\rho}_{2}/dt\\
\widetilde{\rho}_{1}\\
\widetilde{\rho}_{2}
\end{array}\right)\thinspace.
\end{align}

Starting from Eq.\,\eqref{eq:AA}, if we take $\mathbf{x}\sim\exp(-i\omega t),$
then $\lambda=-i\omega$ will be the eigenvalues of the complex matrix
$A,$ and the eigenvalues are determined by the eigen-polynomial of
$A.$ Note that an eigenvalue $\lambda$ and an eigenfrequency $\omega$
are two different quantities. But they are connected by the simple
relation, $\omega=i\lambda.$ In terms of the eigenfrequency $\omega,$
the eigen-polynomial of $A$ is the dispersion relation \eqref{eq:DR}.
The instability diagram determined from the dispersion relation is
shown in Fig.\,\ref{f1}. As discussed in Sec.\,\ref{sec:Introduction},
we will focus on the complex G-Hamiltonian nature of Eq.\,\eqref{eq:AA},
and reveal the fundamental connection between the G-Hamiltonian nature
of Eq.\,\eqref{eq:AA} and the instability structure in Fig.\,\ref{f1}.

\section{G-Hamiltonian system and Krein's theory\label{sec:PB}}

A complex Hamiltonian system \cite{Tekkoyun06} in the space of $(\mathbf{z},\bar{\mathbf{z}})$,
where $\mathbf{z}\in C^{n}$ and $\bar{\mathbf{z}}\in C^{n}$, is
defined by the following Poisson bracket between two functions $f(z,\bar{z})$
and $g(z,\bar{z})$,
\begin{equation}
\{f,g\}=\dfrac{1}{i}\sum_{j}\left(\dfrac{\partial f}{\partial\mathbf{z}_{j}}\dfrac{\partial g}{\partial\mathbf{\bar{z}}_{j}}-\dfrac{\partial f}{\partial\mathbf{\bar{z}}_{j}}\dfrac{\partial g}{\partial\mathbf{z}_{j}}\right),
\end{equation}
and a Hamiltonian function $H(z,\bar{z}).$ The functions $f(z,\bar{z})$,
$g(z,\bar{z})$, and $H(z,\bar{z})$ are complex functions from $C^{n}\times C^{n}$
to $C.$ The dynamical equations for $\mathbf{z}\in C^{n}$ and $\bar{\mathbf{z}}\in C^{n}$
are given by
\begin{align}
\dot{\mathbf{z}} & =\{\mathbf{z},H\}\thinspace,\\
\dot{\bar{\mathbf{z}}} & =\{\bar{\mathbf{z}},H\}\thinspace,
\end{align}
which are explicitly
\begin{align}
\dot{\mathbf{z}} & =\dfrac{1}{i}\dfrac{\partial H}{\partial\bar{\mathbf{z}}}\thinspace,\label{eq:cHz}\\
\dot{\bar{\mathbf{z}}} & =-\dfrac{1}{i}\dfrac{\partial H}{\partial\mathbf{z}}\thinspace.\label{eq:cHzb}
\end{align}

If we require the Hamiltonian function to satisfy the reality condition,
i.e., $H(z,\bar{z}):\thinspace C^{n}\times C^{n}\rightarrow R$ or
\begin{equation}
H(\mathbf{z},\bar{\mathbf{z}})=\overline{H(\mathbf{z},\bar{\mathbf{z}})}\thinspace,\label{eq:rc}
\end{equation}
then it follows that
\begin{equation}
\dfrac{\partial H}{\partial\bar{\mathbf{z}}}=\overline{\dfrac{\partial H}{\partial\mathbf{z}}}\thinspace.\label{eq:rc2}
\end{equation}
To show this, we re-express the Hamiltonian function as $H(\mathbf{z},\bar{\mathbf{z}})=H^{'}(\mathbf{q}(\mathbf{z},\bar{\mathbf{z}}),\mathbf{p}(\mathbf{z},\bar{\mathbf{z}}))\in R$,
where $\mathbf{z}=\dfrac{\mathbf{q}+i\mathbf{p}}{\sqrt{2}}$, and
$\mathbf{q}$ and $\mathbf{p}$ are real. By the chain rule,
\begin{align}
\dfrac{\partial H}{\partial\mathbf{\bar{\mathbf{z}}}} & =\dfrac{\partial H^{'}}{\partial\mathbf{q}}\dfrac{1}{\sqrt{2}}+\dfrac{\partial H^{'}}{\partial\mathbf{p}}(-\dfrac{1}{\sqrt{2}i})\thinspace,\\
\dfrac{\partial H}{\partial\mathbf{z}} & =\dfrac{\partial H^{'}}{\partial\mathbf{q}}\dfrac{1}{\sqrt{2}}+\dfrac{\partial H^{'}}{\partial\mathbf{p}}(\dfrac{1}{\sqrt{2}i})\thinspace,
\end{align}
which implies that Eq.\,\eqref{eq:rc2} holds. As a consequence,
Eqs.\,\eqref{eq:cHz} and \eqref{eq:cHzb} are equivalent. In this
case, the complex Hamiltonian system, i.e., Eqs.\, \eqref{eq:cHz}
and \eqref{eq:cHzb}, are equivalent to the real canonical Hamiltonian
system in terms of $\mathbf{q}$ and $\mathbf{p}$ \cite{Strocchi66},
\begin{equation}
\left(\begin{array}{c}
\dot{\mathbf{q}}\\
\dot{\mathbf{p}}
\end{array}\right)=\left(\begin{array}{cc}
0 & I\\
-I & 0
\end{array}\right)\left(\begin{array}{c}
\partial H^{'}/\partial\mathbf{q}\\
\partial H^{'}/\partial\mathbf{p}
\end{array}\right)\thinspace.
\end{equation}

In the present study, we define a general complex G-Hamiltonian system
for $\mathbf{z}\in C^{n}$ and $\bar{\mathbf{z}}\in C^{n}$ to be

\begin{align}
\dot{\mathbf{z}} & =\dfrac{1}{i}G^{-1}\dfrac{\partial H}{\partial\bar{\mathbf{z}}}\thinspace,\label{eq:GHz}\\
\dot{\bar{\mathbf{z}}} & =-\dfrac{1}{i}\bar{G}^{-1}\dfrac{\partial H}{\partial\mathbf{z}}\thinspace,\label{eq:GHzb}
\end{align}
where $G$ is a non-singular Hermite matrix and the Hamiltonian function
$H(\mathbf{z},\bar{\mathbf{z}})$ satisfies the reality condition
in Eq.\,\eqref{eq:rc}. Because of Eq.\,\eqref{eq:rc2}, Eqs.\,\eqref{eq:GHz}
and \eqref{eq:GHzb} are equivalent, and we only need to investigate
one of them.

For a linear G-Hamiltonian systems satisfying the reality condition,
its Hamiltonian function may assume the form of
\begin{equation}
H(\mathbf{z},\bar{\mathbf{z}})=\mathbf{z}^{T}M\mathbf{z}+\bar{\mathbf{z}}^{T}\bar{M}\bar{\mathbf{z}}+\bar{\mathbf{z}}^{T}S\mathbf{z}\thinspace,\label{eq:GLGH}
\end{equation}
where $M$ is a symmetric matrix and $S$ is a Hermite matrix. In
the present study, we will focus on a special class of linear G-Hamiltonian
system in the form of
\begin{align}
\dot{\mathbf{x}} & =A\mathbf{x}\thinspace,\label{eq:GH}\\
A & =iG^{-1}S\thinspace,\label{eq:A}
\end{align}
where $G$ is a non-singular Hermite matrix and $S$ is a Hermite
matrix. The corresponding Hamiltonian function is
\begin{equation}
H(\mathbf{x})=-\mathbf{x}^{*}S\mathbf{x}\thinspace.\label{eq:SLGH}
\end{equation}
This special linear G-Hamiltonian system is called Hamiltonian system
and has been studied in detail by Yakubovich and Starzhinskii \cite{KGML1958}
without the Hamiltonian structure, i.e., Eq.\,\eqref{eq:GHz} or
\eqref{eq:GHzb}, and the Hamiltonian function \eqref{eq:SLGH}. However,
we will see in Sec.\,\ref{sec:TS} that the Hamiltonian structure
and Hamiltonian function will bring important physical insight to
the mathematical theory of Krein collisions. Since in this paper we
will not discuss nonlinear complex G-Hamiltonian systems and general
linear G-Hamiltonian systems specified by Eq.\,\eqref{eq:GLGH},
the G-Hamiltonian system in the rest of this section and other sections
refers to the special linear G-Hamiltonian system specified by Eqs.\,\eqref{eq:GH}-\eqref{eq:SLGH}.

A matrix $A$ that can be expressed in the form of Eq.\,\eqref{eq:A}
is call a G-Hamiltonian matrix by Yakubovich and Starzhinskii \cite{KGML1958},
and we will follow this convention. An equivalent condition for $A$
to be G-Hamiltonian is that there exists a nonsingular Hermite matrix
$G$ such that
\begin{equation}
A^{*}G+GA=0\thinspace.\label{eq:GH1}
\end{equation}

For any two vectors $\psi$ and $\phi,$ a product is defined as
\begin{equation}
\left\langle \psi,\phi\right\rangle =\phi^{*}G\psi.\label{eq:product}
\end{equation}
Following Yakubovich and Starzhinskii \cite{KGML1958}, we categorize
the eigenvalues of a G-Hamiltonian matrix $A$ according to their
Krein signatures as follows:
\begin{enumerate}
\item For a simple eigenvalue $\lambda$ of $A$ on the imaginary axis,
i.e., $Re(\lambda)=0,$ let the corresponding eigenvector be $\mathbf{y}.$
It can be shown that $\left\langle \mathbf{y},\mathbf{y}\right\rangle \neq0$.
Thus, we define the eigenvalue $\lambda$ to be the first kind if
$\left\langle \mathbf{y},\mathbf{y}\right\rangle >0$, and the second
kind if $\left\langle \mathbf{y},\mathbf{y}\right\rangle <0$.
\item For an $r$-fold eigenvalue $\lambda$ of $A$ on the imaginary axis,
let $V_{\lambda}$ be its eigen-subspace. If $\left\langle \mathbf{y},\mathbf{y}\right\rangle >0$
for any $\mathbf{y}\in V_{\lambda},$ then $\lambda$ is defined to
be the first kind. If $\left\langle \mathbf{y},\mathbf{y}\right\rangle <0$
for any $\mathbf{y}\in V_{\lambda},$ then eigenvalue $\lambda$ is
defined to be the second kind. This category includes (a) as a special
case when $r=1.$
\item For an $r$-fold eigenvalue $\lambda$ of $A$ on the imaginary axis,
let $V_{\lambda}$ be its eigen-subspace. If $\left\langle \mathbf{y},\mathbf{y}\right\rangle =0$
for a $\mathbf{y}\in V_{\lambda},$ then eigenvalue $\lambda$ is
defined to be the mixed kind. An equivalent condition for a mixed
kind is that there exist two eigen-vectors $\mathbf{y}_{1}\in V_{\lambda}$
and $\mathbf{y}_{2}\in V_{\lambda}$, such that $\left\langle \mathbf{y}_{1},\mathbf{y}_{1}\right\rangle >0$
and $\left\langle \mathbf{y}_{2},\mathbf{y}_{2}\right\rangle <0.$
\item For an $r$-fold eigenvalue $\lambda$ of $A$ not on the imaginary
axis, i.e., $Re(\lambda)\neq0,$ the eigenvalue $\lambda$ is defined
to be the first kind if $Re(\lambda)<0,$ and the second kind if $Re(\lambda)>0.$
\end{enumerate}
The first kind is assigned a Krein signature of $+$, the second kind
is assigned a Krein signature of $-$, and the mixed kind is assigned
a Krein signature of $0$. The first kind and the second kind are
also called definite, and the mixed kind is also called indefinite.

Without giving proofs, we list the following theorems regarding the
properties of the eigenvalues of a G-Hamiltonian matrix that will
be used in the present study. The proofs can be found in Ref.\,\cite{KGML1958}.
\begin{thm}
The eigenvalues of a G-Hamiltonian matrix are symmetric with respect
to the imaginary axis.\label{thm:es}
\end{thm}

\begin{thm}
The number of each kind of eigenvalue is determined by the Hermite
matrix $G$. Let $p$ be the number of positive eigenvalues and $q$
be the number of negative eigenvalues of the matrix $G$, then any
G-Hamiltonian matrix has $p$ eigenvalues of first kind and $q$ eigenvalues
of second kind (counting multiplicity). \label{thm:nu}
\end{thm}

\begin{thm}
(Krein-Gel'fand-Lidskii theorem) The G-Hamiltonian system \eqref{eq:GH}
is strongly stable if and only if all of the eigenvalues of $A$ lie
on the imaginary axis and are definite. \label{thm:KGL}
\end{thm}
By definition, a G-Hamiltonian system \eqref{eq:GH} is strongly stable
if all systems nearby are also stable. Systems nearby include those
with system parameters perturbed by an infinitesimal amount relative
to the original system. The Krein-Gel'fand-Lidskii theorem tells us
that for a stable G-Hamiltonina system, the only route for the system
to become unstable when varying the system parameters is through the
overlap between two eigenvalues with different Krein signatures on
the imaginary axis. Such overlaps are called Krein collisions. Once
one of the eigenvalues is ``knocked off'' the imaginary axis, the
system must be unstable, because according to Theorem \ref{thm:es}
the eigenvalues are symmetric with respect to the imaginary axis.

In the next section will we show that the warm two-stream system does
have a G-Hamiltonian structure, and thus its instability is governed
by the Krein collision process.

\section{G-Hamiltonian structure and Krein collision between positive- and
negative- action modes\label{sec:TS}}

It is found that the linear system for the warm two-stream dynamics
described by Eq.\,\eqref{eq:AA} has a G-Hamiltonian structure, because
\begin{equation}
A=\left(\begin{array}{cccc}
-2ikv_{1}^{0} & 0 & (kv_{1}^{0})^{2}-\omega_{p1}^{2}-k^{2}v_{T1}^{2} & -\omega_{p1}^{2}\\
0 & -2ikv_{2}^{0} & -\omega_{p2}^{2} & (kv_{2}^{0})^{2}-\omega_{p2}^{2}-k^{2}v_{T2}^{2}\\
1 & 0 & 0 & 0\\
0 & 1 & 0 & 0
\end{array}\right)=iG^{-1}S,\label{eq:AGS}
\end{equation}
where
\begin{equation}
G=\left(\begin{array}{cccc}
0 & 0 & i\omega_{p2}^{2} & 0\\
0 & 0 & 0 & i\omega_{p1}^{2}\\
-i\omega_{p2}^{2} & 0 & 2kv_{1}^{0}\omega_{p2}^{2} & 0\\
0 & -i\omega_{p1}^{2} & 0 & 2kv_{2}^{0}\omega_{p1}^{2}
\end{array}\right),\label{G}
\end{equation}
and
\begin{equation}
S=\left(\begin{array}{cccc}
\omega_{p2}^{2} & 0 & 0 & 0\\
0 & \omega_{p1}^{2} & 0 & 0\\
0 & 0 & -\omega_{p2}^{2}((kv_{1}^{0})^{2}-\omega_{p1}^{2}-k^{2}v_{T_{1}}^{2}) & \omega_{p1}^{2}\omega_{p2}^{2}\\
0 & 0 & \omega_{p1}^{2}\omega_{p2}^{2} & -\omega_{p1}^{2}((kv_{2}^{0})^{2}-\omega_{p2}^{2}-k^{2}v_{T_{2}}^{2})
\end{array}\right).
\end{equation}
Equation \eqref{eq:AGS} can be verified by direct calculation. According
to Theorem \ref{thm:nu}, the distribution of Krein signatures of
the eigenvalues for $A$ is determined by $G$ given by Eq.\,\eqref{G}.
It has four eigenvalues,
\begin{equation}
\begin{split} & kv_{2}^{0}\omega_{p1}^{2}-\sqrt{(1+(kv_{2}^{0})^{2})\omega_{p1}^{4}},~kv_{2}^{0}\omega_{p1}^{2}+\sqrt{(1+(kv_{2}^{0})^{2})\omega_{p1}^{4}}~,\\
 & kv_{1}^{0}\omega_{p2}^{2}-\sqrt{(1+(kv_{1}^{0})^{2})\omega_{p2}^{4}},~kv_{1}^{0}\omega_{p2}^{2}+\sqrt{(1+(kv_{1}^{0})^{2})\omega_{p2}^{4}}~
\end{split}
\end{equation}
Two of them are positive and the other two are negative. Thus, the
G-Hamiltonian matrix $A$ has two eigenvalues of the first kind and
two eigenvalues of the second kind.

We now show that the band structure of the instability region in Fig.\,\ref{f1}
is produced by the Krein collision process. But, we will first explain
the physical meaning of the Krein signature. For an eigenvalue $\lambda=-i\omega$
of $A$ on the imaginary axis with an eigenvector \textbf{$\mathbf{y}$},
it follows that
\begin{equation}
iG^{-1}S{\bf y}=A\mathbf{y}=-i\omega\mathbf{y}\thinspace.
\end{equation}
Because of Eqs.\,\eqref{eq:SLGH} and \eqref{eq:product}, we obtain
\begin{equation}
\left\langle {\bf y},{\bf y}\right\rangle =\frac{H(\mathbf{y})}{\omega}\thinspace.
\end{equation}
It is the clear that for an eigenvalue $\lambda$ of $A$ on the imaginary
axis, the physical meaning of its signature is the sign of the action,
which is defined to be the ratio between energy and the eigenfrequency
of the mode. This gives us the following physical interpretation of
the celeberated Krein-Gel'fand-Lidskii theorem: the system becomes
unstable when and only when a negative-action mode resonates with
a positive-action mode. We emphasize that it is not accurate to state
that the system becomes unstable when and/or only when a negative-energy
mode resonates with a positive-energy mode, because two positive-energy
modes (or two negative-energy modes) can collide at zero frequency
to destabilize the system. For an eigenvalue $\lambda$ of $A$ that
is not on the imaginary axis, its action is defined to be positive
if $Im(\omega)<0,$ and negative if $Im(\omega)>0.$

\begin{figure}
\includegraphics[scale=0.55]{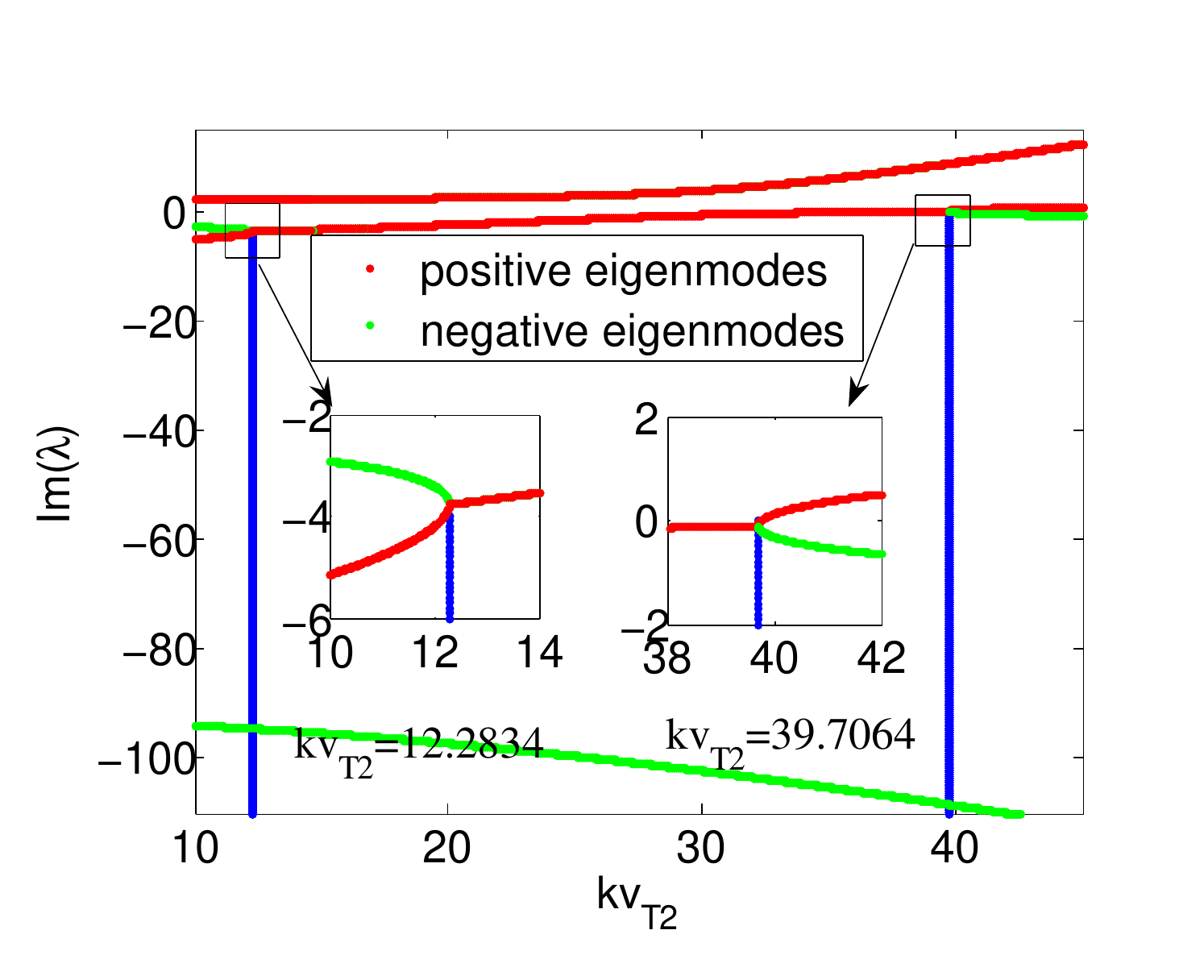}\includegraphics[scale=0.55]{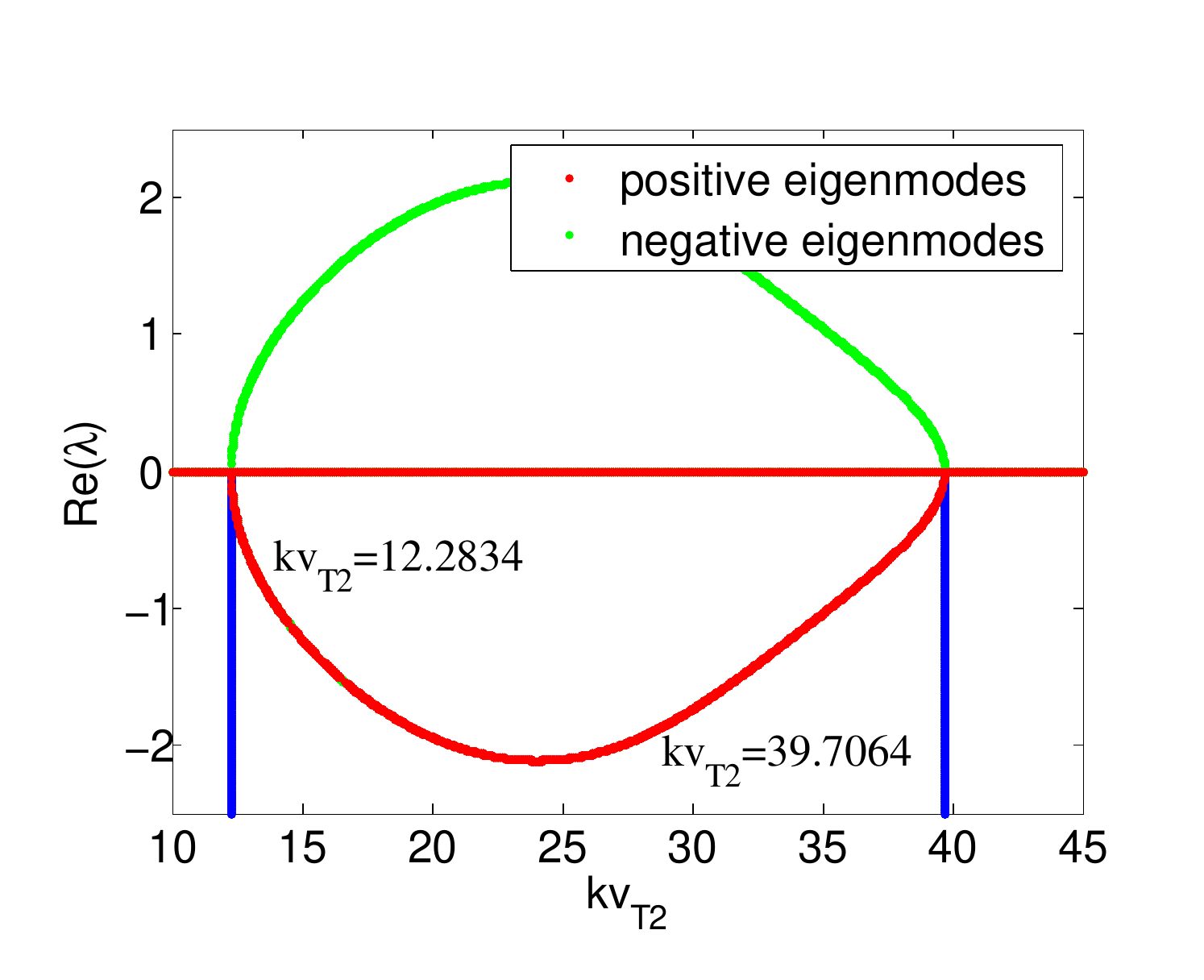}\protect\caption{Plots of Krein collisions when varying $kv_{T_{2}}$ at $w_{p_{1}}^{2}=1,\thinspace w_{p_{2}}^{2}=1836,\thinspace kv_{1}^{0}=0,\thinspace kv_{2}^{60}=50$
and $kv_{T_{1}}=1$. Krein collisions occur at $kv_{T2}=12.2834$
and $kv_{T2}=39.7064$.}
\label{f2}
\end{figure}

Now we show that band of instability region in Fig.\,\ref{f1} is
indeed the result of a Krein collision. In Fig.\,\ref{f2}, we vary
the value of $kv_{T2}$ and fix all other system parameters at $\omega_{p1}^{2}=1,\thinspace\omega_{p2}^{2}=1836,\thinspace kv_{1}^{0}=0,\thinspace kv_{2}^{0}=50$,
and $kv_{T1}=1.$ This motion corresponds to traveling horizontally
in Fig.\,\ref{f1} at $kv_{2}^{0}=50.$ When $kv_{T2}=10$, the eigenvalues
of $A$ are all on the imaginary axis and are distinct, and the system
is stable. Two of the eigenmodes (marked by red) have positive actions
and the other two (marked by green) have negative actions. As $kv_{T2}$
increases, one of the positive-action mode moves towards one of the
negative-action mode. At $kv_{T2}=12.2834$, the two modes with opposite-sign
actions collide and destabilize the system. This point defines the
upper instability boundary in Fig.\,\ref{f1}. When $kv_{T2}$ increases
beyond this threshold, these two eigenvalues move off the imaginary
axis, and the system is unstable. Because the eigenvalues are symmetric
with respect to the imaginary axis according to Theorem \ref{thm:es},
these two complex eigenvalues must have the same imaginary parts,
but opposite real parts. Increasing $kv_{T2}$ to $39.7064$ will
strike the lower instability boundary in Fig.\,\ref{f1}. This corresponds
to another Krein collision between two modes with opposite-sign actions.
When $kv_{T2}>39.7064$, the system will again have four distinct
eigenvalues on the imaginary axis, and the system is stable. There
are two ways to look at the Krein collision at $kv_{T2}=39.7064$.
In the direction of increasing $kv_{T2}$, we observe a Krein collision
between two eigenmodes with opposite-sign actions in an unstable system,
and the system is stabilized by the collision. In the direction of
decreasing $kv_{T2}$, we observe a Krein collision between two stable
eigenmodes with opposite-sign actions, and the system is destabilized
by the collision. Of course, this point-of-view applies to the Krein
collsion at $kv_{T2}=12.2834$ as well.

\begin{figure}
\includegraphics[scale=0.55]{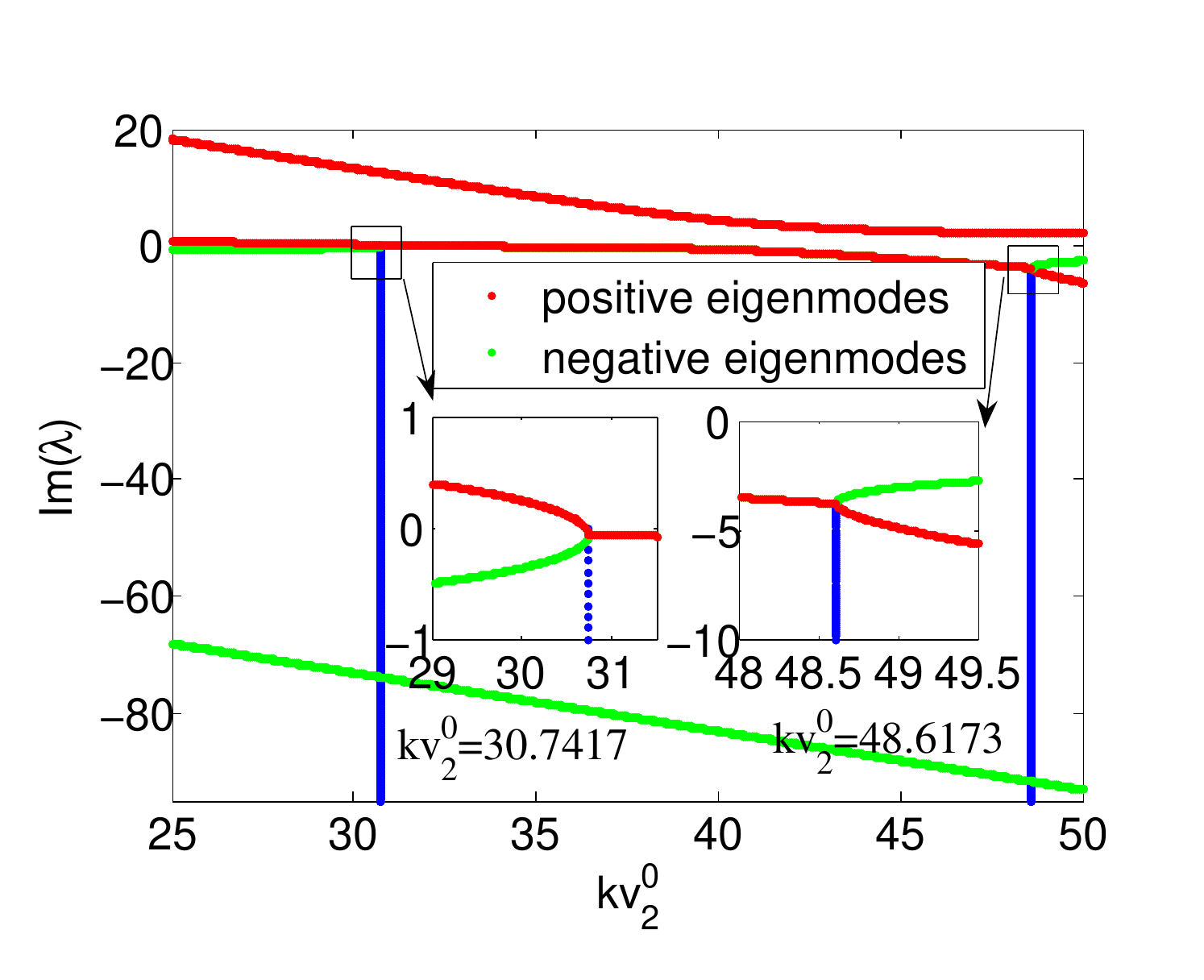}\includegraphics[scale=0.55]{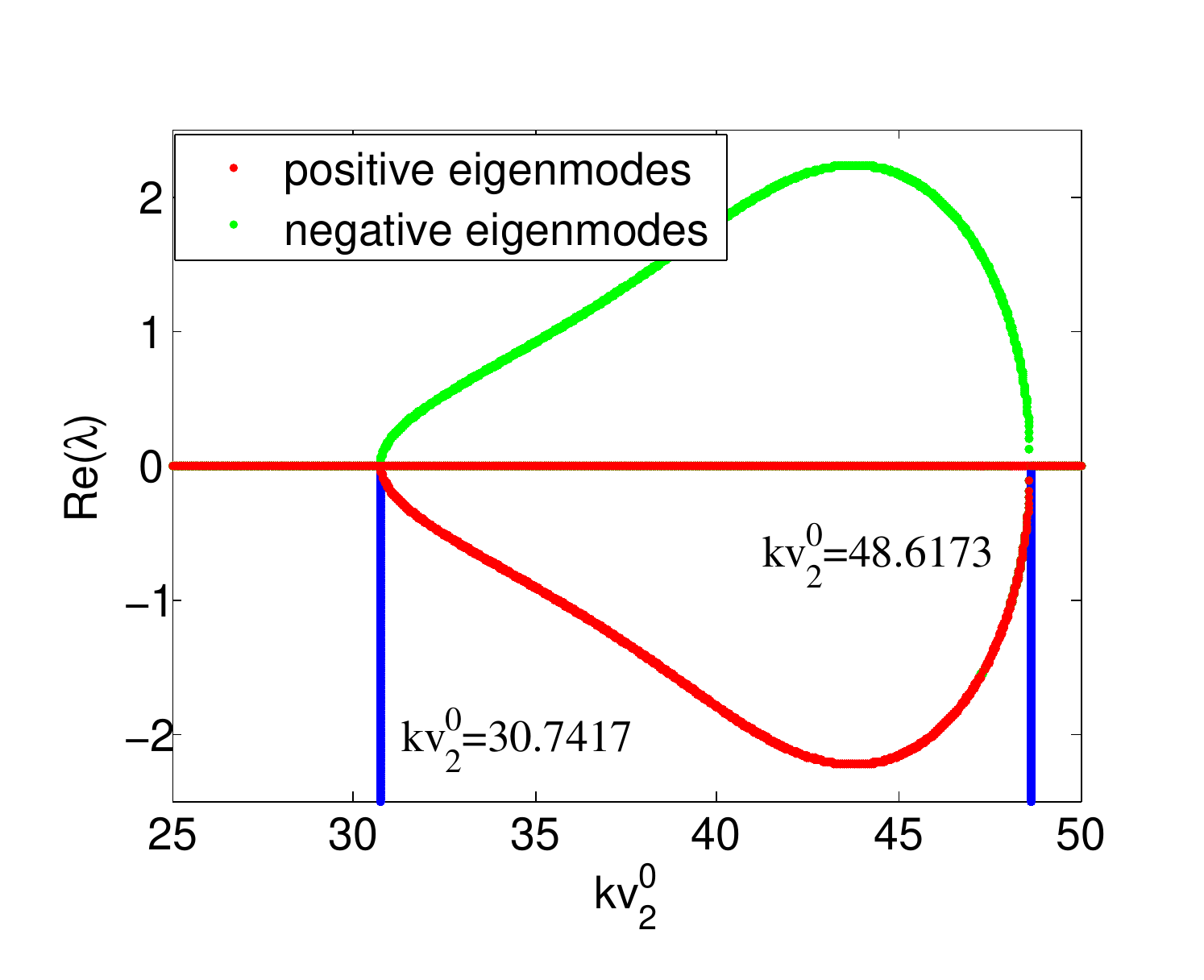}\protect\caption{Plots of Krein collisions when varying $kv_{2}^{0}$ at $w_{p_{1}}^{2}=1,\thinspace w_{p_{2}}^{2}=1836,\thinspace kv_{1}^{0}=0,\thinspace kv_{T1}=1$
and $kv_{T2}=5$. Krein collisions occur at $kv_{2}^{0}=30.7417$
and $kv_{2}^{0}=48.6173$.}
\label{f3}
\end{figure}

In Fig.\,\ref{f3}, we vary $kv_{2}^{0}$ while fixing all other
parameters at $\omega_{p1}^{2}=1,\thinspace\omega_{p2}^{2}=1836,\thinspace kv_{1}^{0}=0,\thinspace kv_{T1}=1,$
and $kv_{T2}=5.$ This corresponds to traveling vertically in Fig.\,\ref{f1}
at $kv_{T2}=5.$ The dynamics of Krein collisions in this motion are
similar to the dynamics in Fig.\,\ref{f2}, i.e., the upper and lower
instability boundaries in this motion correspond to Krein collisions
between two eigenmodes with actions that have opposite signs.

Finally, we describe an interesting phenomenon. We ask the question
whether the system is stable when it is exactly on the boundaries
between the stable and unstable regions, i.e., at the points of the
Krein collisions. From the eigenvalue point of view, the eigenvalues
are on the imaginary axis (or the eigenfrequencies are on the real
axis), one may expect that the system is stable. This is not correct,
because on these boundaries the eigenvalues are repeated eigenvalues.
Whether the system is stable or not depends on whether matrix $A$
can be diagonalized or not. For the warm two-stream system studied
here, it turns out that there is only one Jordan block for the repeated
eigenfreqencies, and the system is algebraically unstable, i.e., the
perturbation grows linearly with time.

\section{Conclusions}

In this paper, we have shown that the dynamics of warm two-stream
modes in wave-number space are governed by a complex G-Hamiltonian
structure, which directly determines the structure of the instability
diagram. We have rigorously shown that the system is destabilized
when and only when a positive action mode resonates with a negative
action mode. It is anticipated that this physical picture of the G-Hamiltonian
structure and destabilization mechanism by resonances between two
modes with actions that have opposite signs holds for other collective
instabilities in conservative systems in plasma physics, accelerator
physics, and fluid dynamics that admit infinite-dimensional Hamiltonian
structures.
\begin{acknowledgments}
This research is supported by the National Natural Science Foundation
of China (NSFC-11505186, 11575185, 11575186), ITER-China Program (2015GB111003,
2014GB124005), the Fundamental Research Funds for the Central Universities
(No. WK2030040068), the China Postdoctoral Science Foundation (No.
2015M581994), the CAS Program for Interdisciplinary Collaboration
Team, the Geo-Algorithmic Plasma Simulator (GAPS) Project, and the
U.S. Department of Energy (DE-AC02-09CH11466).
\end{acknowledgments}
%

\begin{thebibliography}{20}%
\makeatletter
\providecommand \@ifxundefined [1]{%
 \@ifx{#1\undefined}
}%
\providecommand \@ifnum [1]{%
 \ifnum #1\expandafter \@firstoftwo
 \else \expandafter \@secondoftwo
 \fi
}%
\providecommand \@ifx [1]{%
 \ifx #1\expandafter \@firstoftwo
 \else \expandafter \@secondoftwo
 \fi
}%
\providecommand \natexlab [1]{#1}%
\providecommand \enquote  [1]{``#1''}%
\providecommand \bibnamefont  [1]{#1}%
\providecommand \bibfnamefont [1]{#1}%
\providecommand \citenamefont [1]{#1}%
\providecommand \href@noop [0]{\@secondoftwo}%
\providecommand \href [0]{\begingroup \@sanitize@url \@href}%
\providecommand \@href[1]{\@@startlink{#1}\@@href}%
\providecommand \@@href[1]{\endgroup#1\@@endlink}%
\providecommand \@sanitize@url [0]{\catcode `\\12\catcode `\$12\catcode
  `\&12\catcode `\#12\catcode `\^12\catcode `\_12\catcode `\%12\relax}%
\providecommand \@@startlink[1]{}%
\providecommand \@@endlink[0]{}%
\providecommand \url  [0]{\begingroup\@sanitize@url \@url }%
\providecommand \@url [1]{\endgroup\@href {#1}{\urlprefix }}%
\providecommand \urlprefix  [0]{URL }%
\providecommand \Eprint [0]{\href }%
\providecommand \doibase [0]{http://dx.doi.org/}%
\providecommand \selectlanguage [0]{\@gobble}%
\providecommand \bibinfo  [0]{\@secondoftwo}%
\providecommand \bibfield  [0]{\@secondoftwo}%
\providecommand \translation [1]{[#1]}%
\providecommand \BibitemOpen [0]{}%
\providecommand \bibitemStop [0]{}%
\providecommand \bibitemNoStop [0]{.\EOS\space}%
\providecommand \EOS [0]{\spacefactor3000\relax}%
\providecommand \BibitemShut  [1]{\csname bibitem#1\endcsname}%
\let\auto@bib@innerbib\@empty
\bibitem [{\citenamefont {Davidson}\ \emph {et~al.}(1999)\citenamefont
  {Davidson}, \citenamefont {Qin}, \citenamefont {Stoltz},\ and\ \citenamefont
  {Wang}}]{Davidson99}%
  \BibitemOpen
  \bibfield  {author} {\bibinfo {author} {\bibfnamefont {R.~C.}\ \bibnamefont
  {Davidson}}, \bibinfo {author} {\bibfnamefont {H.}~\bibnamefont {Qin}},
  \bibinfo {author} {\bibfnamefont {P.~H.}\ \bibnamefont {Stoltz}}, \ and\
  \bibinfo {author} {\bibfnamefont {T.~S.}\ \bibnamefont {Wang}},\ }\href@noop
  {} {\bibfield  {journal} {\bibinfo  {journal} {Physical Review Special
  Topics-Accelerators and Beams}\ }\textbf {\bibinfo {volume} {2}},\ \bibinfo
  {pages} {054401} (\bibinfo {year} {1999})}\BibitemShut {NoStop}%
\bibitem [{\citenamefont {Davidson}\ and\ \citenamefont
  {Qin}(2001)}]{1davidson2001physics}%
  \BibitemOpen
  \bibfield  {author} {\bibinfo {author} {\bibfnamefont {R.~C.}\ \bibnamefont
  {Davidson}}\ and\ \bibinfo {author} {\bibfnamefont {H.}~\bibnamefont {Qin}},\
  }\href@noop {} {\emph {\bibinfo {title} {Physics of intense charged particle
  beams in high energy accelerators}}}\ (\bibinfo  {publisher} {World
  Scientific},\ \bibinfo {year} {2001})\BibitemShut {NoStop}%
\bibitem [{\citenamefont {Qin}\ \emph {et~al.}(2000)\citenamefont {Qin},
  \citenamefont {Davidson},\ and\ \citenamefont {Lee}}]{qinhong2000}%
  \BibitemOpen
  \bibfield  {author} {\bibinfo {author} {\bibfnamefont {H.}~\bibnamefont
  {Qin}}, \bibinfo {author} {\bibfnamefont {R.~C.}\ \bibnamefont {Davidson}}, \
  and\ \bibinfo {author} {\bibfnamefont {W.~W.}\ \bibnamefont {Lee}},\
  }\href@noop {} {\bibfield  {journal} {\bibinfo  {journal} {Physical Review
  Special Topics-Accelerators and Beams}\ }\textbf {\bibinfo {volume} {3}},\
  \bibinfo {pages} {084401} (\bibinfo {year} {2000})}\BibitemShut {NoStop}%
\bibitem [{\citenamefont {Zimmermann}(2004)}]{zimmermann2004}%
  \BibitemOpen
  \bibfield  {author} {\bibinfo {author} {\bibfnamefont {F.}~\bibnamefont
  {Zimmermann}},\ }\href@noop {} {\bibfield  {journal} {\bibinfo  {journal}
  {Physical Review Special Topic - Accelerators and Beams}\ }\textbf {\bibinfo
  {volume} {7}},\ \bibinfo {pages} {124801} (\bibinfo {year}
  {2004})}\BibitemShut {NoStop}%
\bibitem [{\citenamefont {Qin}\ and\ \citenamefont
  {Davidson}(2014)}]{qin2014two}%
  \BibitemOpen
  \bibfield  {author} {\bibinfo {author} {\bibfnamefont {H.}~\bibnamefont
  {Qin}}\ and\ \bibinfo {author} {\bibfnamefont {R.~C.}\ \bibnamefont
  {Davidson}},\ }\href@noop {} {\bibfield  {journal} {\bibinfo  {journal}
  {Physics of Plasmas}\ }\textbf {\bibinfo {volume} {21}},\ \bibinfo {pages}
  {064505} (\bibinfo {year} {2014})}\BibitemShut {NoStop}%
\bibitem [{\citenamefont {Kueny}\ and\ \citenamefont
  {Morrison}(1995)}]{Kueny95}%
  \BibitemOpen
  \bibfield  {author} {\bibinfo {author} {\bibfnamefont {C.~S.}\ \bibnamefont
  {Kueny}}\ and\ \bibinfo {author} {\bibfnamefont {P.~J.}\ \bibnamefont
  {Morrison}},\ }\href@noop {} {\bibfield  {journal} {\bibinfo  {journal}
  {Physics of Plasmas}\ }\textbf {\bibinfo {volume} {2}},\ \bibinfo {pages}
  {1926} (\bibinfo {year} {1995})}\BibitemShut {NoStop}%
\bibitem [{\citenamefont {Morrison}\ and\ \citenamefont
  {Hagstrom}(2014)}]{Morrison14}%
  \BibitemOpen
  \bibfield  {author} {\bibinfo {author} {\bibfnamefont {P.~J.}\ \bibnamefont
  {Morrison}}\ and\ \bibinfo {author} {\bibfnamefont {G.~I.}\ \bibnamefont
  {Hagstrom}},\ }\enquote {\bibinfo {title} {Nonlinear physical systems šc
  spectral analysis, stability and bifurcations},}\ \ (\bibinfo  {publisher}
  {Wiley},\ \bibinfo {year} {2014})\ Chap.\ \bibinfo {chapter} {12 Continuum
  Hamiltonian Hopf Bifurcation I}\BibitemShut {NoStop}%
\bibitem [{\citenamefont {Lashmore-Davies}(2007)}]{Lashmore07}%
  \BibitemOpen
  \bibfield  {author} {\bibinfo {author} {\bibfnamefont {C.~N.}\ \bibnamefont
  {Lashmore-Davies}},\ }\href
  {http://scitation.aip.org/content/aip/journal/pop/14/9/10.1063/1.2768016}
  {\bibfield  {journal} {\bibinfo  {journal} {Physics of Plasmas}\ }\textbf
  {\bibinfo {volume} {14}},\ \bibinfo {eid} {092101} (\bibinfo {year}
  {2007})}\BibitemShut {NoStop}%
\bibitem [{\citenamefont {Krein}(1950)}]{Krein1950}%
  \BibitemOpen
  \bibfield  {author} {\bibinfo {author} {\bibfnamefont {M.}~\bibnamefont
  {Krein}},\ }\href@noop {} {\bibfield  {journal} {\bibinfo  {journal} {Doklady
  Akad. Nauk. SSSR N.S.}\ }\textbf {\bibinfo {volume} {73}},\ \bibinfo {pages}
  {445} (\bibinfo {year} {1950})}\BibitemShut {NoStop}%
\bibitem [{\citenamefont {Gel'fand}\ and\ \citenamefont
  {Lidskii}(1955)}]{Gel1955}%
  \BibitemOpen
  \bibfield  {author} {\bibinfo {author} {\bibfnamefont {I.~M.}\ \bibnamefont
  {Gel'fand}}\ and\ \bibinfo {author} {\bibfnamefont {V.~B.}\ \bibnamefont
  {Lidskii}},\ }\href@noop {} {\bibfield  {journal} {\bibinfo  {journal}
  {Uspekhi Mat. Nauk}\ }\textbf {\bibinfo {volume} {10}},\ \bibinfo {pages} {3}
  (\bibinfo {year} {1955})}\BibitemShut {NoStop}%
\bibitem [{\citenamefont {Yakubovich}\ and\ \citenamefont
  {Starzhinskii}(1958)}]{KGML1958}%
  \BibitemOpen
  \bibfield  {author} {\bibinfo {author} {\bibfnamefont {V.}~\bibnamefont
  {Yakubovich}}\ and\ \bibinfo {author} {\bibfnamefont {V.}~\bibnamefont
  {Starzhinskii}},\ }\href@noop {} {\emph {\bibinfo {title} {Linear
  Differential Equations with Periodic Coefficients}}},\ Vol.~\bibinfo {volume}
  {I}\ (\bibinfo {year} {1958})\BibitemShut {NoStop}%
\bibitem [{\citenamefont {Sturrock}(1958)}]{Sturrock58}%
  \BibitemOpen
  \bibfield  {author} {\bibinfo {author} {\bibfnamefont {P.}~\bibnamefont
  {Sturrock}},\ }\href@noop {} {\bibfield  {journal} {\bibinfo  {journal}
  {Annals of Physics (N.Y.)}\ }\textbf {\bibinfo {volume} {4}},\ \bibinfo
  {pages} {306} (\bibinfo {year} {1958})}\BibitemShut {NoStop}%
\bibitem [{\citenamefont {Qin}\ and\ \citenamefont
  {Davidson}(2009)}]{Qin09PRSTAB}%
  \BibitemOpen
  \bibfield  {author} {\bibinfo {author} {\bibfnamefont {H.}~\bibnamefont
  {Qin}}\ and\ \bibinfo {author} {\bibfnamefont {R.~C.}\ \bibnamefont
  {Davidson}},\ }\href {\doibase 10.1103/PhysRevSTAB.12.064001} {\bibfield
  {journal} {\bibinfo  {journal} {Phys. Rev. ST Accel. Beams}\ }\textbf
  {\bibinfo {volume} {12}},\ \bibinfo {pages} {064001} (\bibinfo {year}
  {2009})}\BibitemShut {NoStop}%
\bibitem [{\citenamefont {Qin}\ \emph {et~al.}(2013)\citenamefont {Qin},
  \citenamefont {Davidson}, \citenamefont {Chung},\ and\ \citenamefont
  {Burby}}]{PhysRevLett.111.104801}%
  \BibitemOpen
  \bibfield  {author} {\bibinfo {author} {\bibfnamefont {H.}~\bibnamefont
  {Qin}}, \bibinfo {author} {\bibfnamefont {R.~C.}\ \bibnamefont {Davidson}},
  \bibinfo {author} {\bibfnamefont {M.}~\bibnamefont {Chung}}, \ and\ \bibinfo
  {author} {\bibfnamefont {J.~W.}\ \bibnamefont {Burby}},\ }\href {\doibase
  10.1103/PhysRevLett.111.104801} {\bibfield  {journal} {\bibinfo  {journal}
  {Phys. Rev. Lett.}\ }\textbf {\bibinfo {volume} {111}},\ \bibinfo {pages}
  {104801} (\bibinfo {year} {2013})}\BibitemShut {NoStop}%
\bibitem [{\citenamefont {Chung}\ \emph {et~al.}(2013)\citenamefont {Chung},
  \citenamefont {Qin}, \citenamefont {Gilson},\ and\ \citenamefont
  {Davidson}}]{Moses13}%
  \BibitemOpen
  \bibfield  {author} {\bibinfo {author} {\bibfnamefont {M.}~\bibnamefont
  {Chung}}, \bibinfo {author} {\bibfnamefont {H.}~\bibnamefont {Qin}}, \bibinfo
  {author} {\bibfnamefont {E.~P.}\ \bibnamefont {Gilson}}, \ and\ \bibinfo
  {author} {\bibfnamefont {R.~C.}\ \bibnamefont {Davidson}},\ }\href
  {http://scitation.aip.org/content/aip/journal/pop/20/8/10.1063/1.4819830}
  {\bibfield  {journal} {\bibinfo  {journal} {Physics of Plasmas}\ }\textbf
  {\bibinfo {volume} {20}},\ \bibinfo {eid} {083121} (\bibinfo {year}
  {2013})}\BibitemShut {NoStop}%
\bibitem [{\citenamefont {Qin}\ \emph {et~al.}(2014)\citenamefont {Qin},
  \citenamefont {Davidson}, \citenamefont {Burby},\ and\ \citenamefont
  {Chung}}]{Qin14PRSTAB-044001}%
  \BibitemOpen
  \bibfield  {author} {\bibinfo {author} {\bibfnamefont {H.}~\bibnamefont
  {Qin}}, \bibinfo {author} {\bibfnamefont {R.~C.}\ \bibnamefont {Davidson}},
  \bibinfo {author} {\bibfnamefont {J.~W.}\ \bibnamefont {Burby}}, \ and\
  \bibinfo {author} {\bibfnamefont {M.}~\bibnamefont {Chung}},\ }\href
  {\doibase 10.1103/PhysRevSTAB.17.044001} {\bibfield  {journal} {\bibinfo
  {journal} {Phys. Rev. ST Accel. Beams}\ }\textbf {\bibinfo {volume} {17}},\
  \bibinfo {pages} {044001} (\bibinfo {year} {2014})}\BibitemShut {NoStop}%
\bibitem [{\citenamefont {Qin}\ \emph {et~al.}(2015)\citenamefont {Qin},
  \citenamefont {Chung}, \citenamefont {Davidson},\ and\ \citenamefont
  {Burby}}]{Qin15-056702}%
  \BibitemOpen
  \bibfield  {author} {\bibinfo {author} {\bibfnamefont {H.}~\bibnamefont
  {Qin}}, \bibinfo {author} {\bibfnamefont {M.}~\bibnamefont {Chung}}, \bibinfo
  {author} {\bibfnamefont {R.~C.}\ \bibnamefont {Davidson}}, \ and\ \bibinfo
  {author} {\bibfnamefont {J.~W.}\ \bibnamefont {Burby}},\ }\href@noop {}
  {\bibfield  {journal} {\bibinfo  {journal} {Physics of Plasmas}\ }\textbf
  {\bibinfo {volume} {22}},\ \bibinfo {pages} {056702} (\bibinfo {year}
  {2015})}\BibitemShut {NoStop}%
\bibitem [{\citenamefont {Groening}\ \emph {et~al.}(2014)\citenamefont
  {Groening}, \citenamefont {Maier}, \citenamefont {Xiao}, \citenamefont
  {Dahl}, \citenamefont {Gerhard}, \citenamefont {Kester}, \citenamefont
  {Mickat}, \citenamefont {Vormann}, \citenamefont {Vossberg},\ and\
  \citenamefont {Chung}}]{Groening14}%
  \BibitemOpen
  \bibfield  {author} {\bibinfo {author} {\bibfnamefont {L.}~\bibnamefont
  {Groening}}, \bibinfo {author} {\bibfnamefont {M.}~\bibnamefont {Maier}},
  \bibinfo {author} {\bibfnamefont {C.}~\bibnamefont {Xiao}}, \bibinfo {author}
  {\bibfnamefont {L.}~\bibnamefont {Dahl}}, \bibinfo {author} {\bibfnamefont
  {P.}~\bibnamefont {Gerhard}}, \bibinfo {author} {\bibfnamefont {O.~K.}\
  \bibnamefont {Kester}}, \bibinfo {author} {\bibfnamefont {S.}~\bibnamefont
  {Mickat}}, \bibinfo {author} {\bibfnamefont {H.}~\bibnamefont {Vormann}},
  \bibinfo {author} {\bibfnamefont {M.}~\bibnamefont {Vossberg}}, \ and\
  \bibinfo {author} {\bibfnamefont {M.}~\bibnamefont {Chung}},\ }\href
  {\doibase 10.1103/PhysRevLett.113.264802} {\bibfield  {journal} {\bibinfo
  {journal} {Phys. Rev. Lett.}\ }\textbf {\bibinfo {volume} {113}},\ \bibinfo
  {pages} {264802} (\bibinfo {year} {2014})}\BibitemShut {NoStop}%
\bibitem [{\citenamefont {Tekkoyun}\ and\ \citenamefont
  {Cabar}(2006)}]{Tekkoyun06}%
  \BibitemOpen
  \bibfield  {author} {\bibinfo {author} {\bibfnamefont {M.}~\bibnamefont
  {Tekkoyun}}\ and\ \bibinfo {author} {\bibfnamefont {G.}~\bibnamefont
  {Cabar}},\ }\href@noop {} {\bibfield  {journal} {\bibinfo  {journal} {Rend.
  Istit. Mat. Univ. Triest}\ }\textbf {\bibinfo {volume} {XXXVIII}},\ \bibinfo
  {pages} {53} (\bibinfo {year} {2006})}\BibitemShut {NoStop}%
\bibitem [{\citenamefont {Strocchi}(1966)}]{Strocchi66}%
  \BibitemOpen
  \bibfield  {author} {\bibinfo {author} {\bibfnamefont {F.}~\bibnamefont
  {Strocchi}},\ }\href@noop {} {\bibfield  {journal} {\bibinfo  {journal}
  {Reviews of Modern Physics}\ }\textbf {\bibinfo {volume} {38}},\ \bibinfo
  {pages} {36} (\bibinfo {year} {1966})}\BibitemShut {NoStop}%
\end{thebibliography}

%

\end{document}